\documentclass[authoryear,12pt]{elsarticle}

\usepackage{graphicx}
\usepackage{amsmath}
\usepackage{amssymb}
\graphicspath{{figs/}}
\DeclareGraphicsExtensions{.pdf,.jpeg,.png}

\usepackage[caption=false,font=footnotesize,farskip=0em]{subfig}

\usepackage{url}

\usepackage{rotating}
\usepackage{multirow}
\usepackage[margin=1in,footskip=0.25in]{geometry}

\usepackage{siunitx}
\DeclareSIUnit\fahrenheit{\SIUnitSymbolDegree F}
\DeclareSIUnit\mile{mi}
\DeclareSIUnit\foot{ft}
\DeclareSIUnit\voltampere{VA} 
\DeclareSIUnit\perunit{pu} 
\DeclareSIUnit\year{y}
\DeclareSIUnit\var{var} 
\DeclareSIUnit\GWh{GWh}

\usepackage{etoolbox}
\patchcmd{\pprintMaketitle}
  {\hrule\vskip12pt}
  {\hrule\vskip12pt\ifvoid\highlightsbox\else\unvbox\highlightsbox\par\vskip12pt\fi}
  {}{}

\newenvironment{highlights}
  {\global\setbox\highlightsbox=\vbox\bgroup\parindent=0pt }
  {\egroup}
\newsavebox\highlightsbox

\usepackage{textcomp}
\usepackage[textsize = scriptsize,textwidth=22mm]{todonotes}
\usepackage[final]{draftwatermark}
\SetWatermarkLightness{0.9}

\usepackage{etoolbox}
\makeatletter
\patchcmd{\ps@pprintTitle}
  {Preprint submitted}
  {Uploaded to}
  {}{}
\makeatother

\journal{arXiv}

\begin{document}

\begin{frontmatter}

\title{How much energy storage do modern power systems need?}

\author[erg]{Autumn Preskill}
\ead{apetrosgood@gmail.com}
  \author[erg]{Duncan S. Callaway\corref{cor1}}
  \ead{dcal@berkeley.edu}

\cortext[cor1]{Corresponding author}
\address[erg]{Energy and Resources Group, University of California at Berkeley, Berkeley, CA 94720-3050}

\begin{abstract}
The central question we seek to address in this paper is: How rapidly do the operating cost benefits of grid-scale energy storage decline as  installed storage capacity increases?   We use a 240-bus model based on the US Western Interconnection, first optimally locating storage in the network and then dispatching it in a unit commitment model with DC load flow.  The model uses storage to provide frequency regulation, load following, and arbitrage for each hour of a study year, and we investigate a range of scenarios for fuel price and renewables penetration.  We find that value from long-term energy shifting is negligible at all penetrations we investigate, but also that displacing fossil-fueled generators from providing reserves is initially very valuable.  However, in most scenarios the value is negligible beyond 10 GWh of storage, or the equivalent of roughly 6 minutes of average demand in the system.  Above penetrations of 4-8 GWh, storage operating cost benefits are less than estimated capacity values for storage.  We also show that storage has the potential to increase overall carbon emissions in the electricity sector, even when it is not providing significant amounts of arbitrage and is preferentially providing regulation and load following services. 
\end{abstract}


\begin{highlights}
\textit{Highlights}
\begin{itemize}
\item Storage operational value very small at modest penetrations
\item Storage reduces generator starts, but arbitrage value is small
\item Increasing quantities of energy storage increases carbon emissions
\item Carbon emissions increase most in a low gas price / high renewables scenario
\end{itemize}
\end{highlights}


\end{frontmatter}
 
\section{Introduction} 

Energy storage is an important part of a variety of proposed scenarios for future grid evolution~\citep{Williams06012012, Nelson2012, piwko2010western, corbus2009eastern, krajavcic2011planning, jacobson2011providing}. Depending on the type of storage technology, energy storage has the potential to reduce peak loads, decrease the need for conventional ancillary services, postpone infrastructure investment, provide ramp support for renewables, and increase system reliability~\citep{EPRI_2010, eyer_2010}.  At present, energy storage resources are limited, coming primarily from hydropower~\citep{loose_2011}.  In the near term, the buildout of newer technologies -- including compressed air energy storage (CAES), sodium sulfur batteries, fly wheels, and lithium ion batteries -- will depend on the potential for storage to provide value to the current electricity system.

The most obvious use for storage devices is energy arbitrage, where storage devices are charged when prices (or system loads in regions without a real time energy price) are low, and then discharged when prices (or system loads) are high. Several papers have investigated the potential for energy storage to provide arbitrage services in various systems, e.g. \citep{lueken2014effects, hittinger2015bulk, bradbury_2014}. However as the total capacity of storage on the system grows, or if the average spread of energy prices decreases with the addition of high-penetration renewables, the arbitrage value of storage could decline \citep{harris2012unit,sioshansi_storage_2009, hildmann_2011}. Some recent work suggests that revenue from arbitrage does not grow for storage capacities in excess of 5 hours \citep{bradbury_2014}. 

As an alternative, several papers investigate the potential for storage to provide reserves.    \citet{drury_2011}~find that providing reserves in addition to arbitrage services can net an additional \$13-51/kW-yr in revenues for storage devices. \cite{harris2012unit} studied the impact of a range of storage technologies in a unit commitment model that included reserve constraints, and although they did not directly report on the value of providing reserves at different penetration levels, they did find that the operating benefits of storage technologies other than CAES do not justify their economic costs.  \citet{walawalkar_2007} and \citet{sioshansi_value_2010} also demonstrate that there are benefits to providing reserves services using storage devices. Additionally, \citet{byrne_2012} show that regulation is the more valuable service to provide over arbitrage. In combination with Bradbury et. al.'s exploration of the relationship between revenue from arbitrage and duration, these results strongly suggest that optimizing storage device deployment for reserves over arbitrage will maximally add value.

\citet{denholm_value_2009} also raise the possibility that there is value to colocating storage with intermittent renewables, but while they demonstrate the value for individual producers by increasing the proportion of their power they are able to sell over a constrained network, they also indicate that the overall system benefits of storage are maximized when storage is operating according to system-wide price signals, rather than according to an individual generator's needs. Nevertheless, the location of storage is still important in a congested network, as the congestion relief provided by storage can allow power to flow from remote locations to demand centers.


However, these papers shed little light on how these relationships will evolve as we install increasing quantities of storage on the grid.  Services valuable at low penetrations of storage are likely to be less valuable on the margin at higher penetrations.  This paper builds on these earlier investigations by examining how rapidly the operating cost benefits -- including reserve provision \textit{and} arbitrage -- of grid-scale energy storage decline as the quantity installed increases. To answer this question we simulate storage operation over a year at different levels of renewable generation, fuel prices and quantities of storage added.   We use a model that iteratively adds storage capacity to a 240-bus model, choosing buses that maximize locational value. The model then uses the added resources to provide frequency regulation, load following, and arbitrage for each hour of the year.  The choice of which services to provide is determined endogenously, based on the most valuable actions and the current operating constraints of the system.  The model is based on a previously published model of the US Western Interconnection (WECC)~\citep{price_2011}, however our central objective is to make broad conclusions about how storage value depends on a variety of factors, rather than to precisely capture the specific value of storage in the WECC context.  

As with earlier studies that include reserves, we find that storage value does not come from multi-hour energy shifting, but instead from displacing fossil-fueled generators from providing {reserves}; as such systems with higher requirements for regulation and load following provide more opportunity for storage value.  However we find that the operating cost benefits due to storage decline very quickly with increasing penetrations: In most scenarios the value is negligible beyond 10 GWh of storage, or the equivalent of roughly 6 minutes of average demand in the system.  We also find that, above penetrations of 4-8 GWh, storage operating cost benefits are less than hypothetical capacity values for storage -- this suggests that capacity value will dominate investment decisions above those penetrations.  In the long run, if penetrations of wind and solar grow even further and power system fuel mixes evolve to accommodate those changes, energy storage could play a more important role in diurnal (or longer) energy shifting, as in~\cite{mileva_2014}.  However our results suggest that in the short run energy storage has relatively little value at the transmission level.  

While operating cost benefits are small, we do show that storage has the potential to \textit{increase} overall carbon emissions in the electricity sector, even when it is not providing significant amounts of arbitrage and is preferentially providing regulation and load following services.  These carbon increases continue beyond the point where storage provides significant operating cost benefits, meaning that even if the primary economic driver for building storage is capacity value, its day-to-day operations could be detrimental to system-wide carbon emissions.  The carbon emissions increase is greatest in a high renewables / low gas price scenario, which is consistent with the most likely future conditions in light of energy futures prices and renewables capacity expansion rates.

\section{Methods}
\subsection{Model and Solution Method}
The model used for this analysis is an hourly unit commitment model of the Western  Interconnection (also referred to as the Western Electricity Coordinating Council, or WECC).  The model is formulated as a mixed-integer linear program that minimizes system operating costs subject to constraints on generator operation, storage device operation, DC power flow, and reserve requirements, which include both a short-duration regulation service, and a longer-duration load-following service. It is solved using a branch-and-cut algorithm that is implemented using the CPLEX 12.5 C++ library. Values and data sources for any constants in the following section are described in more detail in Section \ref{sec:data_inputs}.

\subsubsection{Objective Function}
The objective function minimizes total system operating costs over the set $\mathcal{G}$ of generators on the system and the set of time periods modeled, $\mathcal{T}$, as follows:
\begin{equation}\min  \,\, \sum_{g \in \mathcal{G}} \sum_{t \in \mathcal{T}} \Gamma_{g} q_{gt} + SU_{g} s_{gt}, \end{equation}
where $q_{gt}$ is a decision variable denoting the level of output for generator $g$ in time period $t$ and $s_{gt}$ is a decision variable denoting whether or not  generator $g$ started up in hour $t$.  
Both $\Gamma_{g}$, the marginal cost for running generator $g$, and $SU_{g}$, the startup cost for generator $g$, depend on the fuel price $F_g$ for generator $g$. Explicitly, $\Gamma_{g} = F_{g} * HR_{g} + O_{g}$, where $HR_g$ and $O_{g}$ are respectively the heat rate for generator $g$ and the variable operations and maintenance cost for generator $g$.\footnote{For simplicity we assume heat rate is constant across each generator's output range} We define startup cost as $SU_{g} = SE_{g} * F_{g} + SA_g$, where $SE_g$ is the energy required to start generator $g$, and $SA_g$ is the fixed cost component of starting generator $g$.

\subsubsection{Generator Constraints}
In each time period, each generator $g$ in the model can supply regulation up, $r^{u}_{gt}$ and regulation down, $r^{d}_{gt}$ as well as load following up, $lf^{u}_{gt}$ and load following down, $lf^{d}_{gt}$, in addition to the energy supplied, $q_{gt}$. These terms are related by the following two constraints:
\begin{align}
q_{gt} + r^{u}_{gt} + lf^{u}_{gt} & \leq  \overline{Q}_{g} u_{gt} && \forall g \in \mathcal{G}, t \in \mathcal{T}, \\
q_{gt} - r^{d}_{gt} - lf^{d}_{gt} & \geq  \underline{Q}_{g} u_{gt} &&\forall g \in \mathcal{G}, t \in \mathcal{T},
\end{align}
where $u_{gt}$ is a binary decision variable denoting whether or not generator $g$ is operating in time period $t$, and $\overline{Q}_{g}$ and $\underline{Q}_{g}$ are the maximum and minimum generation limits, respectively, for generator $g$. Each of the ancillary service variables must also be less than their respective limits for each generator:
\begin{alignat}{4}
0  & \leq r^{u}_{gt}  && \leq RU_{g}u_{gt}  &&\,\,\,\,\,\,\,\,\,\,\,\,&\forall g \in G, t \in T\\
0  & \leq r^{d}_{gt}  &&\leq RD_{g}u_{gt} &&\,\,\,\,&\forall g \in \mathcal{G}, t \in \mathcal{T}\\
0  & \leq lf^{u}_{gt}  && \leq LFU_{g}u_{gt} &&\,\,\,\,&\forall g \in \mathcal{G}, t \in \mathcal{T}\\
0  & \leq lf^{d}_{gt}  && \leq LFD_{g}u_{gt} &&\,\,\,\,&\forall g \in \mathcal{G}, t \in \mathcal{T}
\end{alignat}
Between hours, generators are subject to ramp rate constraints:
\begin{align} 
R^-_{g} &\leq q_{gt} - q_{g,t-1} - r^{d}_{gt} - lf^{d}_{gt} & \forall g \in \mathcal{G}, t \in \mathcal{T} \\
R^+_{g} &\geq q_{gt}-q_{g,t-1} + r^{u}_{gt} + lf^{u}_{gt}& \forall g \in \mathcal{G}, t \in \mathcal{T}
\end{align}
Continuous startup variables for generators are used with binary operating variables and minimum up and down times in the manner described by \citet{papavasiliou_2011}:
\begin{alignat}{2}
\sum_{k=t-UT_{g}+1}^{t} s_{gk} & \leq u_{gt} \,\,\,\,\,\,\,\,&\forall g \in \mathcal{G}, t \in \mathcal{T}\\
 \sum_{k=t+1}^{t+DT_{g}} s_{gk} & \leq 1 - u_{gt} \,\,\,\,\,\,\,\,&g \in \mathcal{G}, t \in \mathcal{T}
 \end{alignat}
 \begin{align}
  s_{gt}  & \geq  u_{gt}  -  u_{g,t-1} \,\,\,\,\,\,&\forall g \in \mathcal{G}, t \in \mathcal{T}\\
 0  & \leq s_{gt}  \leq  1 \,\,\,\,\,\,&\forall g \in \mathcal{G}, t \in \mathcal{T}\\
 u_{gt} & \in \{0,1\} \,\,\,\,\,\,\,&\forall g \in \mathcal{G}, t \in \mathcal{T}.
 \end{align}

\subsubsection{Storage Constraints}
We model energy arbitrage in each storage device as scheduled consumption or supply of energy in hourly blocks.  We also model the commitment of storage capacity to provide regulation and load following reserves and enforce constraints on reserves that avoid ``double counting'' capacity, i.e. if capacity is committed to providing reserves it cannot also be used for arbitrage. 

Energy in storage device $m$ at time $t$, $e_{mt}$ must be less than the capacity $E_m$ of the storage device, where $\mathcal{M}$ is the set of all storage devices on the system:
\begin{equation} 0 \leq e_{mt} \leq E_{m} \,\,\,\,\,\forall m \in \mathcal{M}, t \in \mathcal{T} \end{equation}
The charge and discharge rates for the storage device are also constrained by the power limits ($P^{\rm charge}$, $P^{\rm discharge}$) of the storage device:
\begin{align}
0 \leq c_{mt} &\leq P^{\rm charge}_{m} & \forall m \in \mathcal{M}, t \in \mathcal{T} \\
0 \leq d_{mt} &\leq P^{\rm discharge}_{m} & \forall m \in \mathcal{M}, t \in \mathcal{T} \\
\end{align}
All storage devices must also satisfy energy balance for scheduled arbitrage in all hours, such that
\begin{equation} e_{mt} = e_{m,t-1} + \tau \beta_m c_{mt} -  \frac{\tau}{\delta_m} d_{mt}, \label{eqn:storage_energy_balance}\end{equation} 
where $c_{mt}$ and $d_{mt}$ are scheduled hourly charge and discharge rates, respectively, between hour $t-1$ and hour $t$, 
$\tau$ is the time period length in hours, and $\beta$ and $\delta$ are the charging efficiency and discharging efficiency, respectively, of storage device $m$.

We enforce constraints to ensure that each battery is capable of serving the worst case reserve action in addition to delivering or consuming energy according to the arbitrage schedule.  Discharging, regulation up, and load following up all require energy to leave the storage device, so the sum of the energy needed to provide each of those services (in the worst case where reserves are provided at the full contracted power for the total contracted duration) in hour $t$ must be greater than the amount available at the beginning of the hour. Similarly, charging, regulation down, and load following down rely on headroom in the storage device, so their sum must be less than the head room available at the beginning of the hour. These constraints are represented as follows:
\begin{align}
e_{m,t-1} & \geq \frac{1}{\delta_m} \left( \tau d_{mt} + \tau^{r} r^{us}_{mt} + \tau^{lf} lf^{us}_{mt} \right), \label{eqn:storage_reserve_up}\\
E_m - e_{m,t-1} & \geq  \beta_m \left( \tau c_{mt} + \tau^{r} r^{ds}_{mt} + \tau^{lf} lf^{ds}_{mt} \right),\label{eqn:storage_reserve_down}
\end{align}
where $\tau^{r}$ is the length of time for which regulation must be provided in hours, $\tau^{lf}$ is the length of time for which load following must be provided in hours, and $r^{us}_{mt}$, $r^{ds}_{mt}$, $lf^{us}_{mt}$, and $lf^{ds}_{mt}$  represent the power contributions of the storage device at node $n$ to regulation up, regulation down, load following up, and load following down, respectively, in time period $t$ (we will define these parameters in Section~\ref{sec:data_inputs}).

\subsubsection{Network Constraints}
We enforce nodal power balance constraints for hourly schedules with a linear DC load flow model:
\begin{equation}\sum_{g \in G_n}(q_{gt}) + \sum_{m \in M_n} (c_{mt} -  d_{mt})  + \sum_{i \in N} B_{ni} ( \theta_{nt}-\theta_{it})  =  L_{nt}, \label{eqn:node_balance}\end{equation}
where $G_n$ is the subset of generators located at node $n$, $M_n$ is the subset of generators located at node $n$, $B_{ni}$ is the susceptance between node $n$ and node $i$, $\theta_{nt}$ is the voltage angle at node $n$ at time $t$, and $L_{nt}$ is the load at node $n$ at time $t$.

Also, the total load flow on line $ij$ must be less than or equal to the maximum load flow allowed, $\overline{D}_{ij}$: 
 \begin{equation}B_{ij} ( \theta_{it}-\theta_{jt}) \leq \overline{D}_{ij}  \end{equation}
 
Note that we do not model power flow associated with reserve actions; we assume that any line capacity violations that result from reserve actions are sufficiently small or short in duration that they can be tolerated by the system operator or, in the case of larger disturbances, that the system can be redispatched to resolve constraints.  We assume these events are sufficiently rare that they can be neglected for our objective of quantifying the annual cost benefits of storage at the scale of the model. 

\subsubsection{Reserve Requirements}
In each hour, minimum reserves of each type (regulation in up and down directions, load following in up and down directions) must be procured, corresponding to system needs. We model daily regulation up and down requirements as a proportion, $\rho$, of the peak load for the day added to a proportion, $\sigma$, of the total installed wind and solar capacity. We model load following up for each hour as a proportion, $\eta$, of the forecasted load plus a proportion, $\nu$, of the forecasted wind and solar for the hour.   We model the load following down requirement as a constant proportion of the renewables forecast.  The following equations define these constraints explicitly, with $\overline{S}_n$ and $\overline{W}_n$ being the solar and wind capacities installed at node $n$, respectively, and $S_{nt}$ and $W_{nt}$ being the solar and wind forecasts at node $n$ during time period $t$. To reduce complexity, we model total reserves constraints globally. 
\begin{align}
& \sum_{g \in G} (r^{u}_{gt})+\sum_{m \in M} (r^{us}_{mt}) \geq \rho \left(\max_{a \in T: t_{max} - a \geq t} L_{na}\right) + \sigma \left( \sum_{n \in N}(\overline{S}_n + \overline{W}_n )\right)& \forall t \in \mathcal{T} \\
  & \sum_{g \in G} (r^{d}_{gt})+\sum_{m \in M} (r^{ds}_{mt}) \geq \rho \left(\max_{a \in T: t_{max} - a \geq t} L_{na}\right) + \sigma \left(\sum_{n \in N}(\overline{S}_n + \overline{W}_n )\right)& \forall t \in \mathcal{T} \\
  & \sum_{g \in G} (lf^{u}_{gt})+\sum_{m \in M} (lf^{us}_{mt}) \geq  \eta \sum_{n\ in N }L_{nt} + \nu \sum_{n \in N}(S_{nt} + W_{nt}) & \forall t \in \mathcal{T} \\
  & \sum_{g \in G} (lf^{d}_{gt})+\sum_{m \in M} (lf^{ds}_{mt}) \geq \nu \sum_{n \in N}(S_{nt} + W_{nt}) & \forall t \in \mathcal{T} 
\end{align}

\subsubsection{Solution Method} We run the model iteratively for each of the days in a given year, passing the final storage levels, generator output levels for ramping, and generator operating and starting levels as constants denoting the starting levels for the next day. This corresponds to the following constraints, where the $prev$ superscript denotes variables from the previous day's solve:
\begin{align}
e_{n0} &= e_{n24}^{prev} &\forall g \in G \\
u_{gb} &= u_{g,24+b}^{prev} &\forall g \in G, b \in (-DT_g+1, ... , 0) \\
s_{gb} &= s_{g,24+b}^{prev} &\forall g \in G, b \in (\min(-UT_g+1, -DT_g+1), ... , 0)
\end{align}Additionally, because it would otherwise be optimal to fully discharge storage devices at the end of each unit commitment modeling period, we also constrain the final storage levels and generator operating levels. To do this, we run a preliminary two-day unit commitment model with a four hour time step for the generator unit commitment variables, and save the generator and storage states at the end of the first day for use as constraints in a second run. In the second (final) run, we use single-day unit commitment in one hour increments with final storage charge levels and final generator operating states constrained to be equal to those saved from the first run (as in \citep{sioshansi_value_2010}). This corresponds to the following additional constraints for the first two-day unit commitment, where $T = \{ t \in \mathbb{Z}: 1 \leq t \leq 48\}$
\begin{equation} u_{gt} = u_{g,t-1} = u_{g,t-2} = u_{g,t-3} \,\,\forall g \in G, \{t \in T: t \mod{4} = 0\} \end{equation}

We implement the model in C++ and solve it with CPLEX 12.5. We solve the first two-day unit commitment problem with a mip gap of 0.5\%, and the second problem with a mip gap of 0.05\%. The average time taken to solve these two problems and obtain results for an individual day was 72.4 seconds. 

\subsection{Data Inputs \label{sec:data_inputs}}
\begin{figure}
\begin{center}
\includegraphics[trim=7cm 10cm 7cm 10cm]{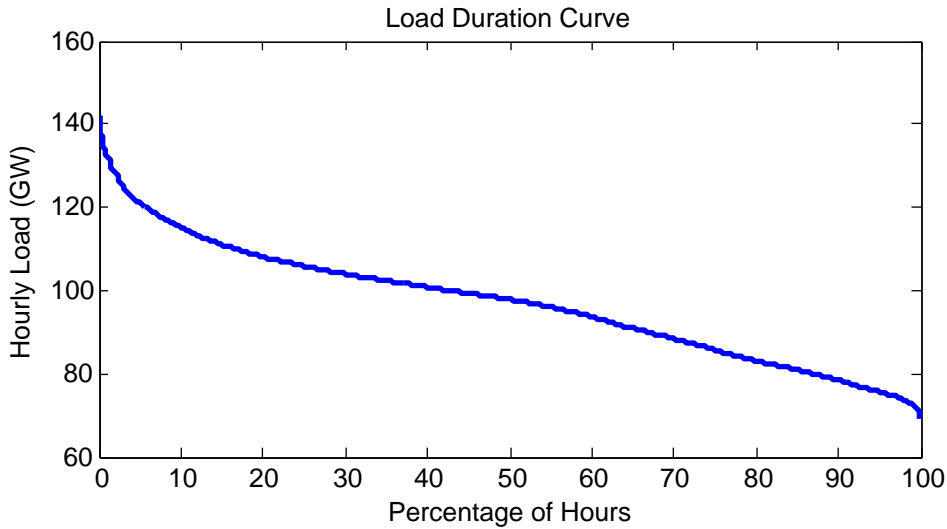}
\caption{The load duration curve for the year of data modeled. \label{fig:loadduration}}
\end{center}
\end {figure}

The layout of the system network for the model is based on data for the 240-bus model created and published in association with a model developed at CAISO~\citep{price_2011}, hereafter the Price model.   From this resource, we obtain susceptances $B_{ij}$ and line limits $\overline{D}_{ij}$  for the network. The hourly load at each node, $L_{nt}$, also comes from the Price model, and is based on 2004 data.  Though WECC infrastructure has evolved since that time\footnote{In the time since the model was built, total demand has remained relatively flat~\citep{wecc_2013som} and generation capacity for all fuels but wind, solar and natural gas were virtually unchanged~\citep{EIA_electricitygen}.  Gas capacity has grown significantly since 2004, however because total and peak demand remained flat this capacity has had relatively little impact on operations.  We will discuss wind and solar additions later in the paper.}, 
our objective in using the Price model is to capture the effect of storage operating over large scales, but not to precisely model the effect of storage on current infrastructure.  However as we will discuss, we will investigate how storage additions impact system operations in different renewables penetration scenarios.  Figure~\ref{fig:loadduration} shows the yearly load duration curve.

\begin{figure}
\begin{center}
\includegraphics[trim=7cm 10cm 7cm 10cm]{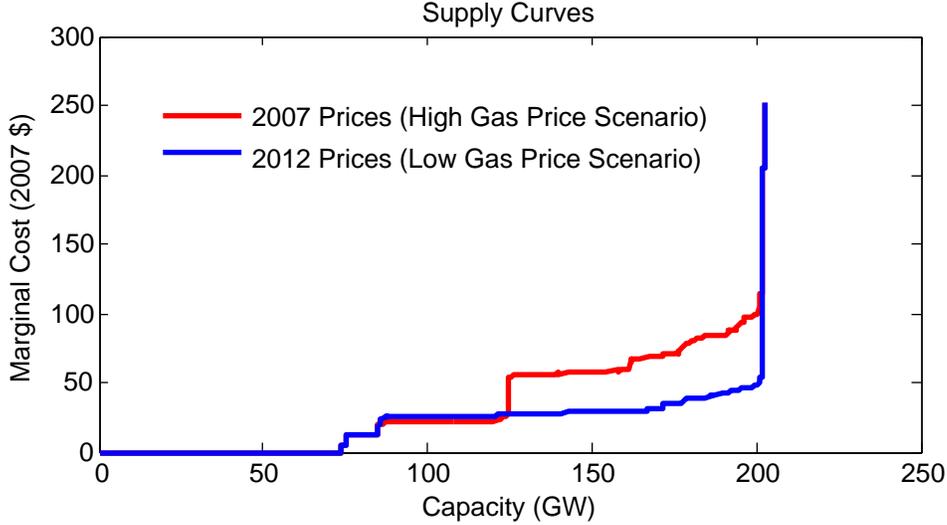}
\caption{The supply curves for the set of generators modeled, in the low and high gas price scenarios.  The supply curves in this graph are created for the low renewables scenarios; the high renewables scenarios have a larger region where the marginal cost is zero.  \label{fig:supply_curve}}
\end{center}
\end {figure}

In total, the model commits and dispatches 185 generators, of which 38 are coal-fired, 135 are gas-fired, 4 are nuclear, and 8 are run on fuel oil.   The model does not dispatch hydro, biomass, wind, solar, and geothermal plants; instead the production profiles and capacities for those generators originate in the Price model. The set of dispatched generators used is based on disaggregated generator data from the Price model, which are then modified such that generators with similar heat rates are aggregated together, and each node in the network has only one generator with each heat rate, which reduces symmetry in the subsequent formulation.  From the Price data we obtain heat rates and maximum operating capacities for each generator ($HR_g$, $\overline{Q}_g$). We obtain fuel prices $F_g$ from EIA data corresponding to 2007 and 2013 \citep{EIA_2013}. Figure~\ref{fig:supply_curve} shows how the heat rate curves for all generators with non-zero marginal cost on the system change with fuel prices. 

We match the prime mover for a generator in the Price data, when available, to TEPPC generator category data from the 2009 TEPPC Study Program Results to obtain ramp limits ($R_{g}^+$, $R_{g}^-$), minimum up- and down-times ($UT_g$, $DT_g$), minimum operating capacities ($\underline{Q}_g$), start-up costs and startup energy required ($SA_g$, $SE_g$), and variable operations and maintenance costs ($O_g$).  When only a fuel type, rather than a prime mover is available from the Price data, we chose the generator type from the TEPPC data with the heat rate that is the closest to the heat rate reported from the data in the Price model and use the corresponding figures.

In each hour, we enforce ancillary service constraints for regulation and load following. We model these on the requirements used in \citet{piwko2010western}. Total regulation in both directions must be greater than 1\% of peak load ($\rho=0.01$) in both directions. The Western Wind Integration Study indicates that 1\% of peak is acceptable for regulation with respect to wind capacity, but does not investigate whether this also applies for additions of solar. To ensure that regulation needs are satisfied with the addition of both resources, we also add 1\% of the installed wind and solar capacities to the regulation requirement in both directions ($\sigma=0.01$).
Total load following in the up direction must be greater than the sum of 3\% of forecasted load and 5\% of forecasted wind and solar ($\eta=0.03$, $\nu=0.05$), in accordance with the "3+5" rule. In accordance with the need for load following in the down direction as specified in~\citet{makarov_operational_2009}, we also require an amount of reserve in the down direction equal to 5\% of forecasted wind and solar. 

The maximum regulation ($RU_g$, $RD_g$) and load following capabilities ($LFU_g$, $LFD_g$) of each generator are calculated based on the maximum generator movement in 10 minutes, using the one-minute ramp rate for the generator's prime mover~\citep{TEPPC_2009}.  Generator limits on ramps between hours were calculated based on maximum generator movement in 60 mins.~\citep{papavasiliou_2011} 

In addition to the generators, the model also dispatches 4 pumped-hydro plants in all scenarios.  The efficiencies and capabilities for the pumped hydro plants are taken from the Price model, and comprise 3.0 GW of power, with 201 GWh of total energy capacity. 

We assume that storage efficiency is 90\% on both charge ($\beta_n$) and discharge ($\delta_n$) and a power:energy ratio of 4, such that $P^{discharge}_{n} / E_n = 4$. By choosing this ratio, we ensure that power constraints will bind for regulation, and energy constraints will bind for load following and arbitrage.  Both pumped hydro and added storage can provide regulation and load-following, subject to constraints that require enough energy to be present in the battery (or energy capacity for charging in the case of down reserves) for provision of 15 minutes of regulation and 2 hours of load following (Eq.~\eqref{eqn:storage_reserve_up} and Eq.~\eqref{eqn:storage_reserve_down} with $\tau^r$ = 0.25 hrs and $\tau^l$ = 2 hrs) \citep{sioshansi_value_2010}.

\subsection{Placing Storage in the Model}

We determined the locations for storage devices prior to running the unit-commitment model by slightly modifying the model.  Specifically, we include decision variables denoting the total amount of energy storage capacity to be added at each node and, for each total storage quantity $E_{\rm tot}$ we investigate, we constrain the sum of storage capacity across all nodes such that $\sum_{\forall n} E_n\le E_{\rm tot}$.  Because these added decision variables significantly increase the complexity of the model we made several modifications to limit computing time in the placement phase.  First, we only run the model on the peak demand day. Second, because reserves are not a location-specific quantity in the model, we dropped reserve requirements from the objective function.  Finally, we identified storage locations iteratively, i.e. after locating the smallest quantity of storage, we fix its location and identify the location of the next increment of storage, and so on.  Because of these modifications to the model, added increments of storage are only optimal for energy arbitrage on the peak demand day.  However, because the peak demand day is the most severely constrained and energy is the only quantity subject to nodal balance constraints, we assume that the identified locations are a decent proxy for the true optimal locations. In all scenarios, storage is preferentially located in San Diego before any other locations.

\subsection{Scenarios}

We explored four different scenarios that allowed us to explore the effects on the value of storage of high vs. low natural gas prices and high vs. low penetrations of renewables.  First, we explored low and high natural gas prices.  We used average fuel prices from recent years in which gas prices were relatively high (2007; the ``high gas price" scenario, \$7.12/MMBtu for gas and \$1.77/MMBtu for coal, 2007 \$ \citep{EIA_2008}) and relatively low (2012; the ``low gas price" scenario, \$3.17/MMBtu for gas and \$2.22/MMBtu for coal \citep{EIA_2013}).    Figure~\ref{fig:supply_curve} shows a supply curve for the generators and prices modeled. For each natural gas price, we also looked at a low-penetration renewables scenario and a high-penetration renewables scenario that meets California's RPS goals \citep{price_2011}.  The low penetration scenario has 6.5 GW of wind and 0.5 GW of solar, whereas the high-penetration scenario has 24 GW of wind, and 7 GW of solar.\footnote{These scenarios are taken directly from the Price model; wind and solar capacity in WECC in the most recent available year (2013) were approximately 18 and 5 GW, respectively~\citep{EIA_electricitygen}}  We also perform a low wind, high solar scenario high-penetration scenario where we scale solar profiles and wind profiles such that their respective installed capacities are reversed.

 \section{Results}

\subsection{Total Social Benefit from Storage}
\begin{figure}
\begin{center}
\includegraphics[trim=7cm 9.5cm 7cm 11cm]{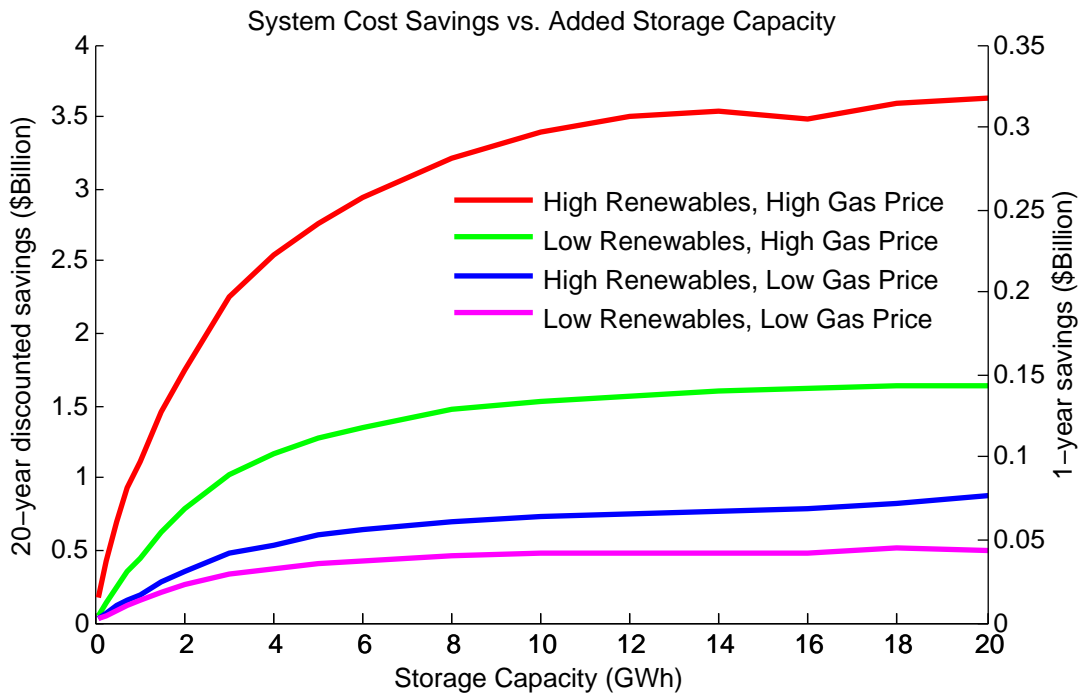}
\caption{System cost savings as storage penetration is increased.  System cost savings level out in each scenario, and by the time 20 GWh of additional storage are added, increasing the amount of storage on the system no longer produces significant savings. 
 \label{fig:savings_storage_capacity}}
\end{center}
\end {figure}

\begin{figure}
\begin{center}
\includegraphics[trim=7cm 10.5cm 7cm 11cm]{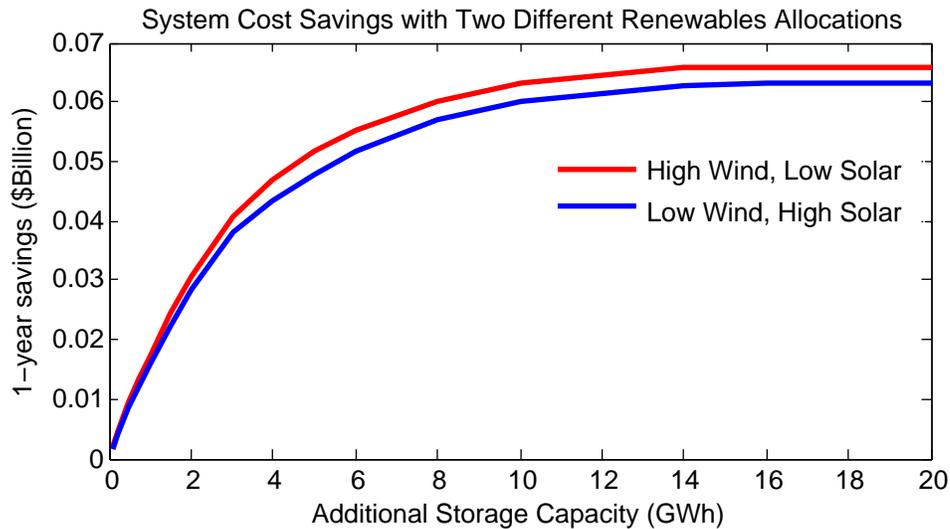}
\caption{In this graphic, the renewable energy provided in the high renewables / low gas price scenario has been reallocated, such that about 66\% of the total renewable energy comes from solar (the low wind / high solar scenario). The default case has been provided as the high wind / low solar scenario. The distribution of energy between wind and solar does not significantly affect the value provided by storage. }
\label{fig:renewable_energy_allocations}
\end{center}
\end {figure}

Figure~\ref{fig:savings_storage_capacity} shows the total system cost savings for each of the four major scenarios as a function of  storage energy added to the system. The range of benefits across scenarios is very large, and both renewables penetration levels and fuel prices have significant impact on the outcome, though for the scenarios we investigate fuel prices appear to matter more.  Note that, because some capacity is allocated to regulation (requiring a 4C battery) and some to spinning reserve (requiring a 1/4C battery), we report results only in units of energy storage (GWh) on the x-axis, rather than units of power capacity. We also find that, for a given penetration of renewables, the specific mix of solar and wind has little impact (less than 4\%) on savings attributable to storage (Figure~\ref{fig:renewable_energy_allocations}).  This is in contrast to other recent results that suggest storage is more important in systems with high solar versus high wind penetration (e.g.~\citep{mileva_2014}).  We attribute this difference to several observations: (1) our model optimizes only the \textit{operation} of the system but not the mix of generation infrastructure as in~\citep{mileva_2014} and (2) storage does relatively little net load shifting in our model and instead is allocated to the higher value ancillary services; we will discuss this more when we describe Fig.~\ref{fig:ReserveBenefits} below.  Finally, we note that storage will be more important for load shifting at higher solar penetrations. 

\begin{figure}
\begin{center}
\includegraphics[trim=7cm 9cm 7cm 11cm]{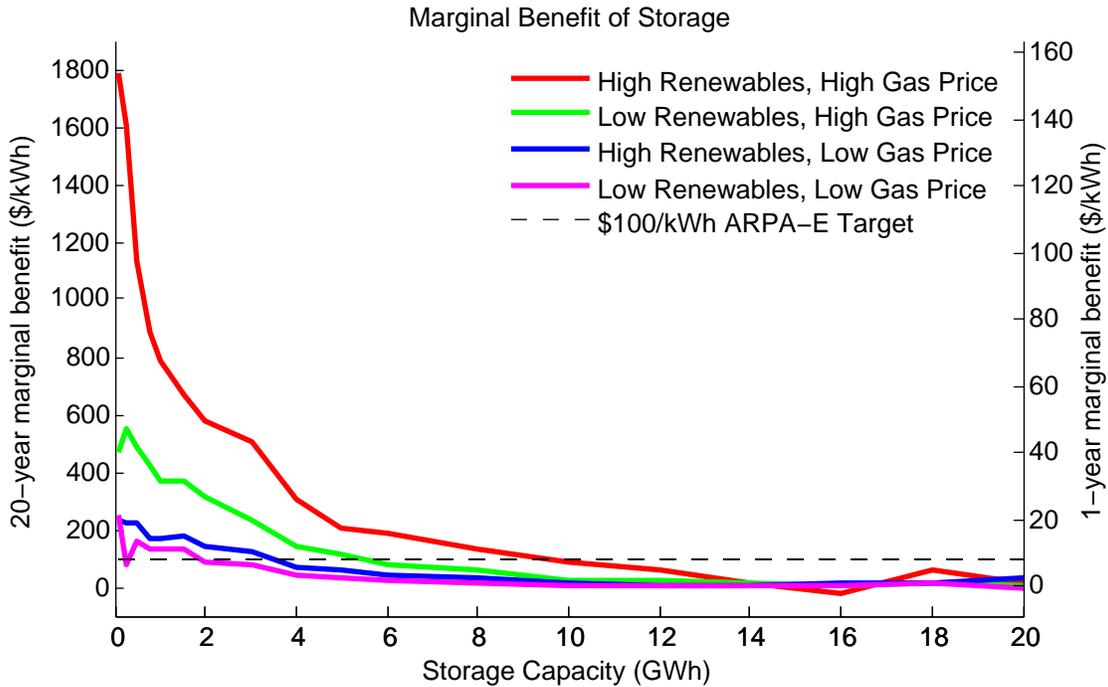}
\caption{Marginal benefit of additional storage. In all scenarios, the benefit of adding an additional unit of storage decreases as the total amount of storage capacity on the system is increased. \label{fig:marginal_cost}}
\end{center}
\end {figure}

Figure~\ref{fig:marginal_cost} shows the marginal benefit for storage for a 20-year time horizon with a 7\% discount rate, assuming identical cost savings each year (left vertical axis), and also for a single year (right vertical axis). The marginal benefit is computed as the ratio of the change in operating cost resulting from each incremental addition of storage to the size of the storage increment. We see that the diversity in value of a small addition of storage capacity across scenarios is large, with the 20-year benefit ranging from \$1800/kWh in the high gas price / high renewables scenario to about \$200/kWh in both low gas price scenarios. Still, with small amounts of storage on the system, the marginal benefit of additional storage in every scenario is greater than the ARPA-E GRIDS target of \$100/kWh~\citep{ARPA-E_2010} (depicted as a dashed line in Figure~\ref{fig:marginal_cost}),  suggesting that discounted system benefits would be greater than storage capital costs for storage technologies that meet this target.\footnote{ARPA-E's target is for systems 1C systems, i.e. batteries that discharge their rated energy in 1 hour.  Frequency regulation requires higher power rating (we assume 4C, or systems that discharge their rated energy in 1/4 hour), which would add to the cost of the technology, possibly significantly more.  Therefore \$100/kWh in our context should be taken as an especially low and aggressive target.} The marginal benefit then drops off sharply, such that by the time 10 GWh of storage capacity have been added to the system,\footnote{For context, at an average demand of 97.1 GW (the average for the dataset used here), 10 GWh of storage capacity could supply average demand in the model for 6 minutes and 11 seconds.} the marginal benefit in all scenarios falls below \$100/kWh.  By the time 20 GWh of additional storage are added, increasing the amount of storage on the system no longer produces significant operating cost savings.  We note that current battery costs are significantly higher than the ARPA-E target, however we do not consider those costs here because the industry is in a phase of rapid cost reduction and policies to support energy storage are likely to be based on the potential for cost:benefit ratios to be attractive in the future rather than today.  

The most likely scenario for the future is one in which gas prices are low, and in these scenarios the \$100/kWh break-even point allows no more than 4 GWh of energy storage capacity on the system before capital costs (at the \$100/kWh target) are no longer recovered through system benefits. Capital costs depend strongly on technology type and power to energy ratio, and battery-only cost estimates (i.e. not including balance of system costs) currently span a very broad range~\citep{nykvist_2015}, although many are working to develop storage devices that will meet the ARPA-E GRIDS target at scale \citep{zhang_2012,narayanan_2012}.

We can also compare these marginal benefits to the capacity value\footnote{By \textit{capacity value} we mean the amount of power a device can make available during peak load conditions.} that storage might provide.  To do this, we divide the lowest cost of conventional generation capacity (we used the overnight cost for a combustion turbine, taken from~\citep{BV_2012} as \$650/kW) by the number of hours storage would need to operate to be qualified as providing capacity value to the system (we assume 4 hours, based on recent requests for offers in California~\citep{SDGE_storage_RFO}). These parameters give an approximate storage capacity value of \$160/kWh.  In the low gas price scenarios, the marginal value of storage for providing arbitrage and ancillary services quickly falls below this number, suggesting that capacity payments will be an important factor for storage investment in these conditions.  On the other hand, for the high gas price / high renewables scenario, capacity payments may not drive investment in storage until higher penetrations (8-10 GWh across the system).  Note that capacity value will be higher in ``load pockets'' with strong constraints on citing conventional generators (e.g. the Los Angeles basin); analysis of these conditions is outside the scope of this paper.

\begin{figure}
\begin{center}
\includegraphics[trim=4cm 5cm 4cm 7cm scale=0.9]{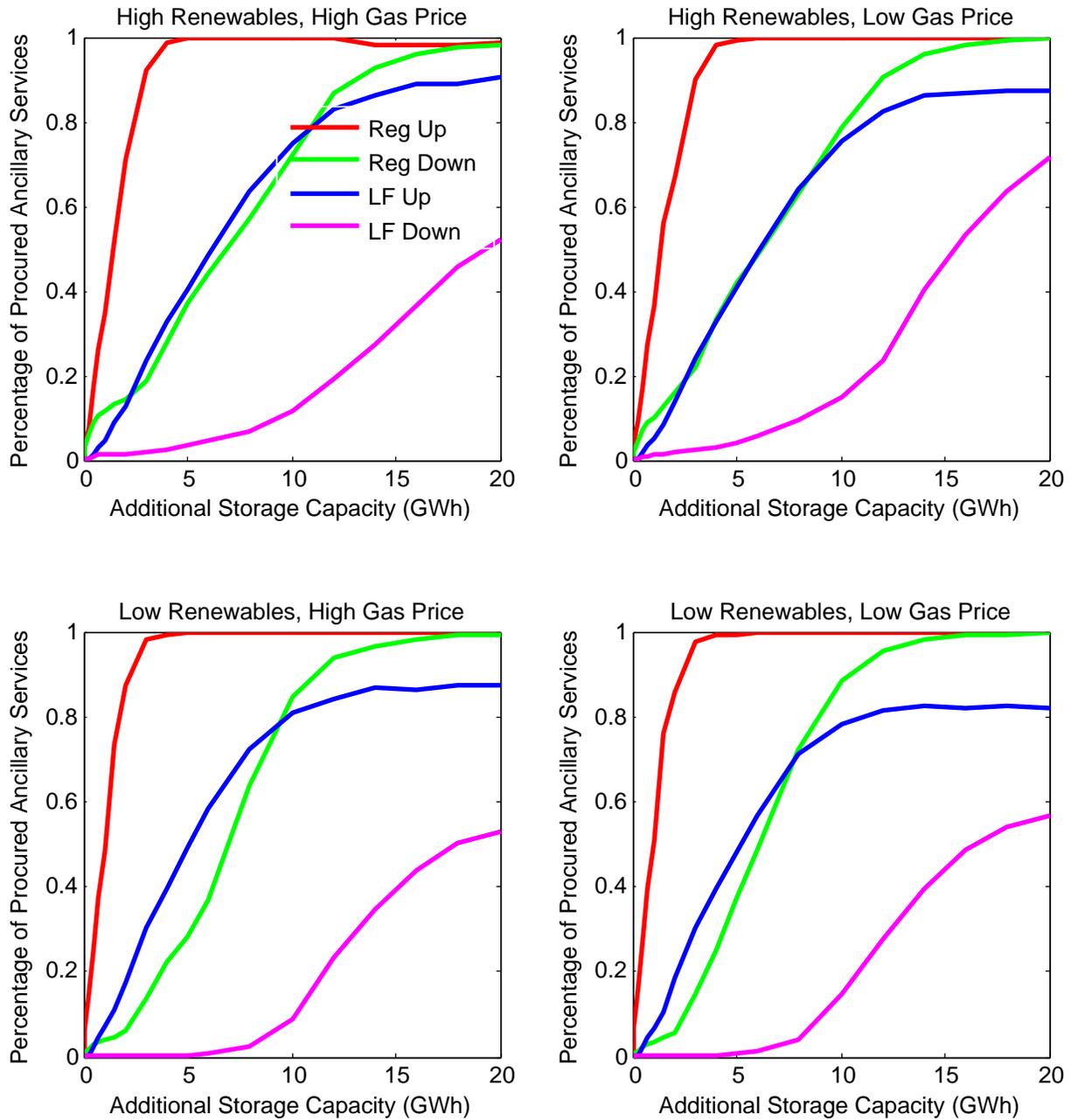}
\caption{Proportions of ancillary services served by storage devices in various scenarios. In all scenarios, storage quickly moves to provide all required regulation up.  Subsequently, storage emphasizes the provision of regulation down and load following up, and then finally begins to increase the proportion of load following down provided. \label{fig:as_proportion}}
\end{center}
\end {figure}

Figure~\ref{fig:as_proportion} shows the proportions of various ancillary services requirements that are satisfied by storage.  Because storage satisfies regulation up requirements first, these results indicate that regulation up is the most valuable service for storage to satisfy.  The next most valuable services are load following up and regulation down, and finally load following down. With a higher gas price, more of the load following up requirement is satisfied by storage. While it might otherwise make sense to satisfy both load following up and regulation down services with the same storage device, in practice this will be undesirable, as regulation requires higher power capacity than load following, and a single storage device will likely be better suited for one or the other service. For investment purposes, it may be reasonable to invest first in high power, lower energy capacity storage devices that will supply regulation up and down, and then later install more moderate power devices with higher energy capacities that can satisfy load following requirements in both directions. 

\begin{figure}
\begin{center}
\includegraphics[trim=7cm 10cm 7cm 11cm]{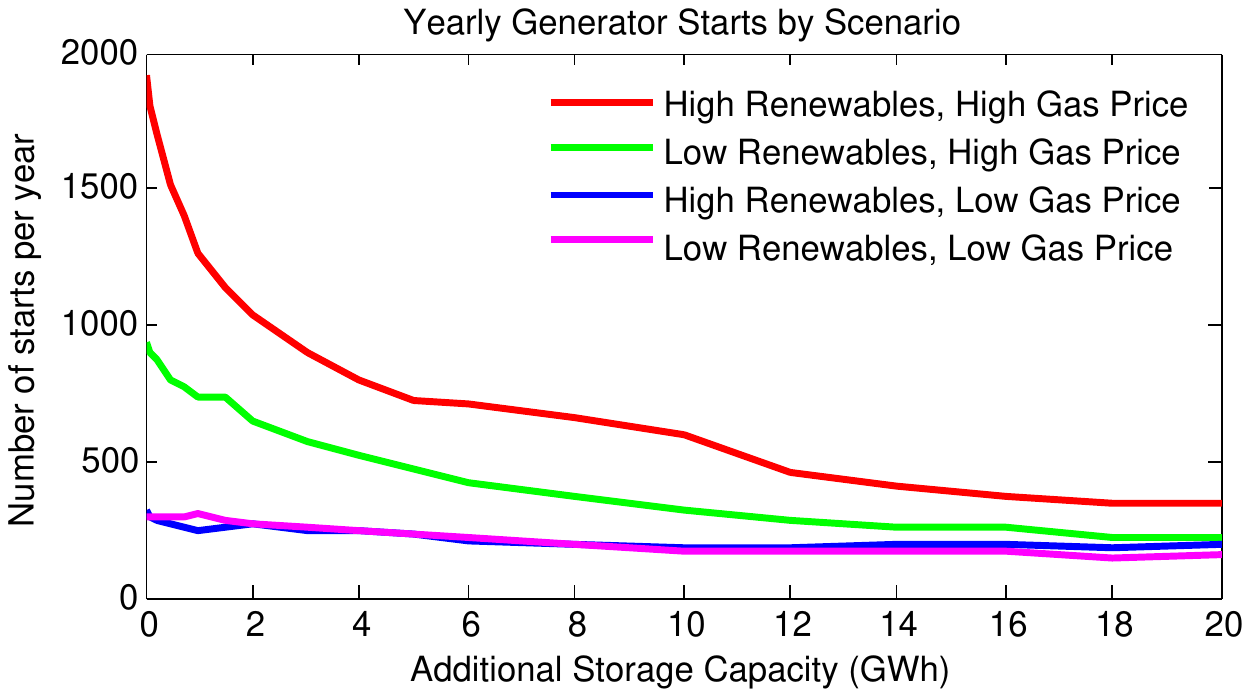}
\caption{Number of generator starts per year.  As the total amount of additional storage capacity on the system increases, the total number of generator starts decreases.  The resulting reduction in startup costs paid contributes to the corresponding decreases in total system operating costs, as shown in Figure~\ref{fig:savings_storage_capacity}. \label{fig:starts}}
\end{center}
\end {figure}

\begin{figure}
\begin{center}
\includegraphics[trim=7cm 10cm 7cm 11cm]{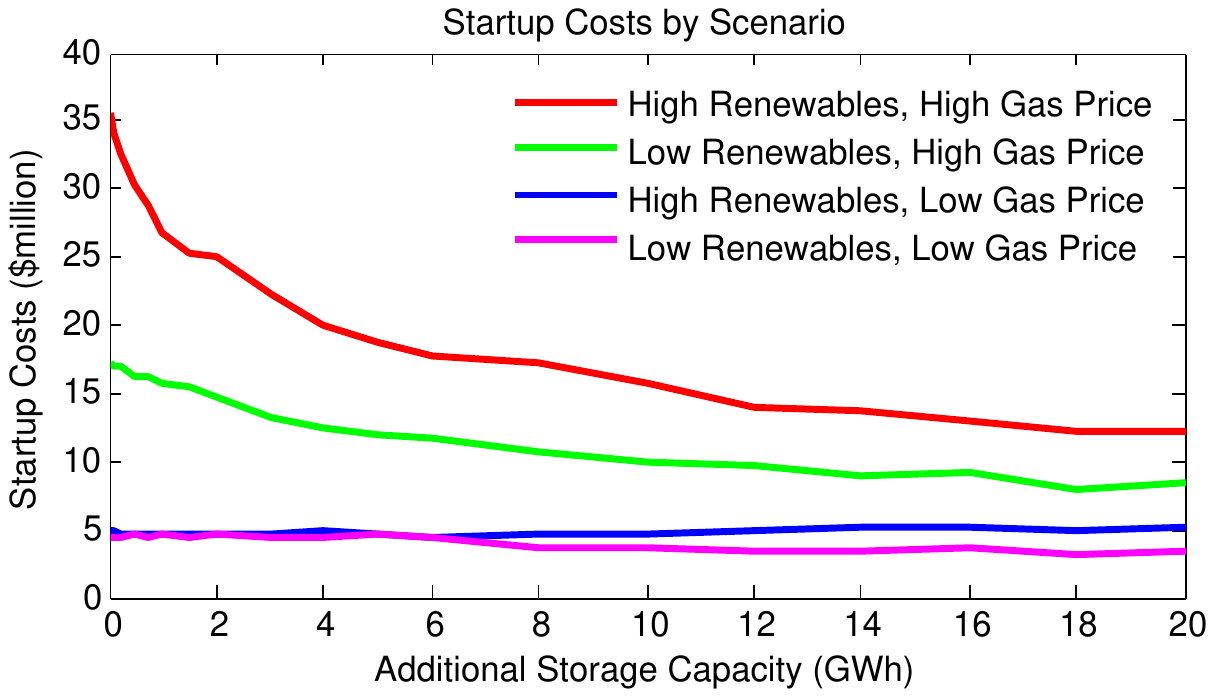}
\caption{Cost of generator starts.  As the total amount of additional storage capacity on the system increases, the total cost of generator starts decreases.  The resulting reduction in startup costs paid contributes to the corresponding decreases in total system operating costs, as shown in Figure~\ref{fig:savings_storage_capacity}. \label{fig:startcosts}}
\end{center}
\end {figure}

Figures~\ref{fig:starts} and~\ref{fig:startcosts} indicate that the number of generator starts and the cost due to generator starts both decrease in all scenarios as the amount of storage present in the system is increased.  The savings due to reductions in generator starts is roughly 10 percent in the most impactful scenario (high renewables penetrations and high gas prices). With lower gas prices, the reduction in generator starts due to storage devices is much less dramatic than in the scenarios with higher gas prices. This is likely due to the fact that, in lower gas price scenarios, generators need to be turned off for a longer length of time to make incurring their startup costs economical.

\subsection{Private and Market Benefits from Storage}
In this section we investigate storage device profits and whether the system benefit from storage can be captured by independent storage operators. For energy market revenue, we assume each generator or storage device is paid the locational marginal price (LMP) for the node at which it is located.   We obtain this price from the dual of the node balance constraint Eq.~\eqref{eqn:node_balance}, which we will call $\lambda_{nt}$, where $n \in \mathcal{N}$ and $t \in \mathcal{T}$.  Assuming a competitive market, we compute the market clearing price for each reserve market in each hour as the maximum opportunity cost (\$/MW) faced by a generator that is providing the corresponding resource in that hour. We will refer to these hourly prices as $\lambda^{ru}_t$, $\lambda^{rd}_t$, $\lambda^{lfu}_t$, and $\lambda^{lfd}_t$ for regulation up, regulation down, load following up, and load following down, respectively. Only generators constrained by their maximum capacities (for generators providing up reserves) or minimum capacities (for generators providing down reserves) experience opportunity costs.  Generators that have not committed their full, currently available capacities are indifferent to committing their capacities to one market versus another; they have available capacity to do both \citep{wu_2004}.  

The gross profit, $z_i$, for a given storage device $i$ over the entire year, then, is calculated as follows:
\begin{equation}
z_i = \sum_{t \in T} \left(\lambda_{it}d_{it} - \lambda_{it}c_{it} + \lambda^{ru}_{t}r_{it}^{us} + \lambda^{rd}_{t}r_{it}^{ds} + \lambda^{lfu}_{t}lf_{it}^{us} + \lambda^{lfd}_{t}lf_{it}^{ds}\right)
\end{equation}
The total gross profit in the system, $Z$, is the sum of the $z_i$'s over all storage devices in the system ($i \in \mathcal{S}$).   Gross profit is calculated as the revenue received in the energy, regulation, and load following markets, less the cost to charge storage with energy purchased in the market.  We do not include other costs or revenues in this metric (for example storage capital costs, taxes and depreciation, or tax incentives).

\begin{figure}
\begin{center}
\includegraphics[trim=4cm 6cm 4cm 7cm]{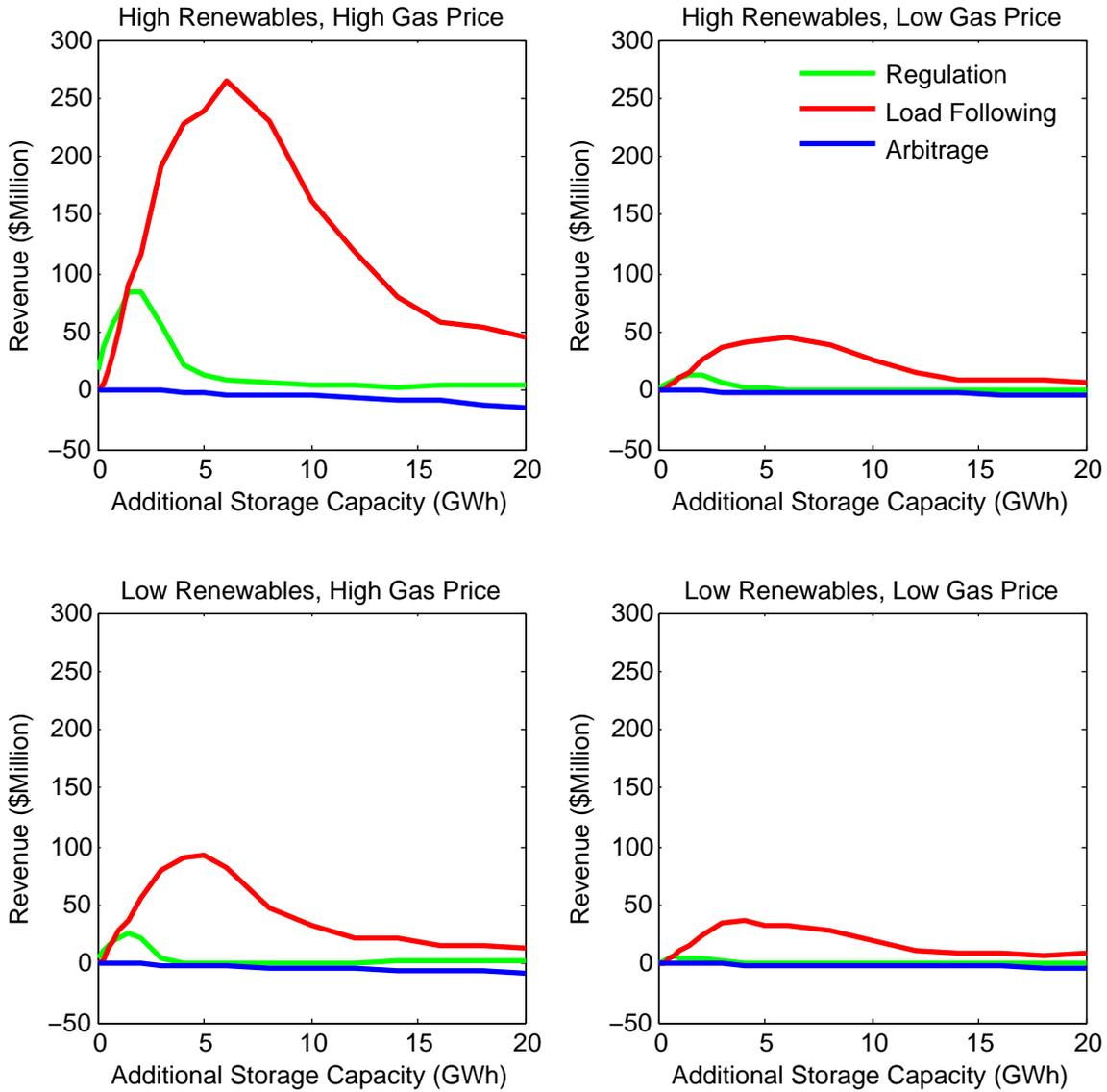}
\caption{Revenue obtained by storage due to each service provided. Revenue is calculated by paying the storage devices the market clearing price for each service provided.  In general, as storage is added to the system, the total revenue achieved by storage decreases. At very high penetrations, storage devices lose money in energy markets so that they may participate in the ancillary services markets.   \label{fig:ReserveBenefits}}
\end{center}
\end {figure}

Figure~\ref{fig:ReserveBenefits} shows the changes in the value of the contribution to $Z$ of reserves, load following, and energy arbitrage as additional storage devices are added to the system.  The total revenue available is largest in the high renewables / high gas price case, when the reserve requirements are the largest due to the renewables, and the market clearing prices are set by generators with higher marginal fuel costs.  
The value to storage operators is coming from reserves more than arbitrage; in fact, as the total amount of storage on the system increases, storage operators lose money in the energy markets in favor of making capacity available for the more lucrative regulation and load following markets. Notably, while the regulation markets are the most lucrative initially, the revenue in these markets drops off quickly, and load following is the service that provides the most revenue over the largest range of installed storage capacities.

\begin{figure}
\begin{center}
\includegraphics[trim=7cm 10cm 7cm 10cm]{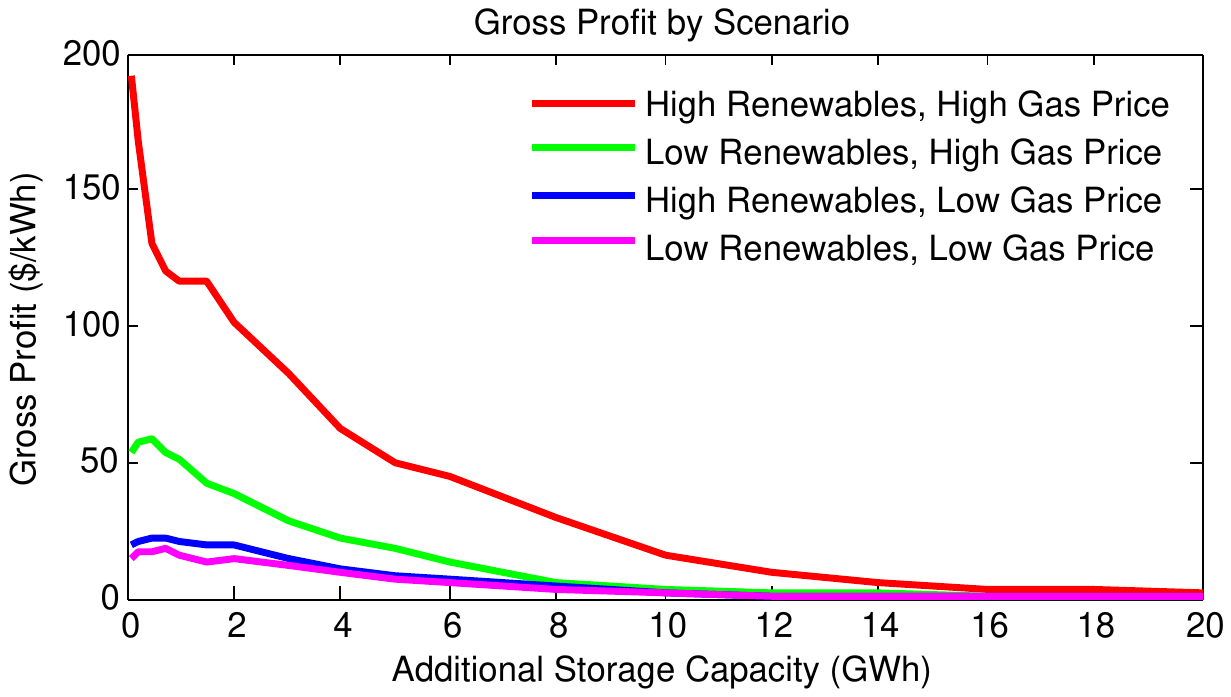}
\caption{Gross profit, calculated as revenues less costs to charge, per kWh installed. 
As storage is added to the system, the gross profit seen by all storage operators decreases.    \label{fig:GrossProfit}}
\end{center}
\end {figure}

Figure~\ref{fig:GrossProfit} shows the marginal changes in $Z$ as the total amount of storage in the system is increased.  The total market revenue is largest in the high renewables / high gas price case, when the reserve requirements are the largest due to the renewables, and the market clearing prices are set by generators with higher marginal fuel costs due to the higher gas prices. Similar to the operating cost benefits, this metric also declines rapidly, and once the system has at least 10 GWh of capacity installed, the market revenue for energy and ancillary services available to storage operators becomes small and unlikely to cover storage capital costs on their own.

\begin{figure}
\begin{center}
\includegraphics[trim=7cm 9.5cm 7cm 11.5cm]{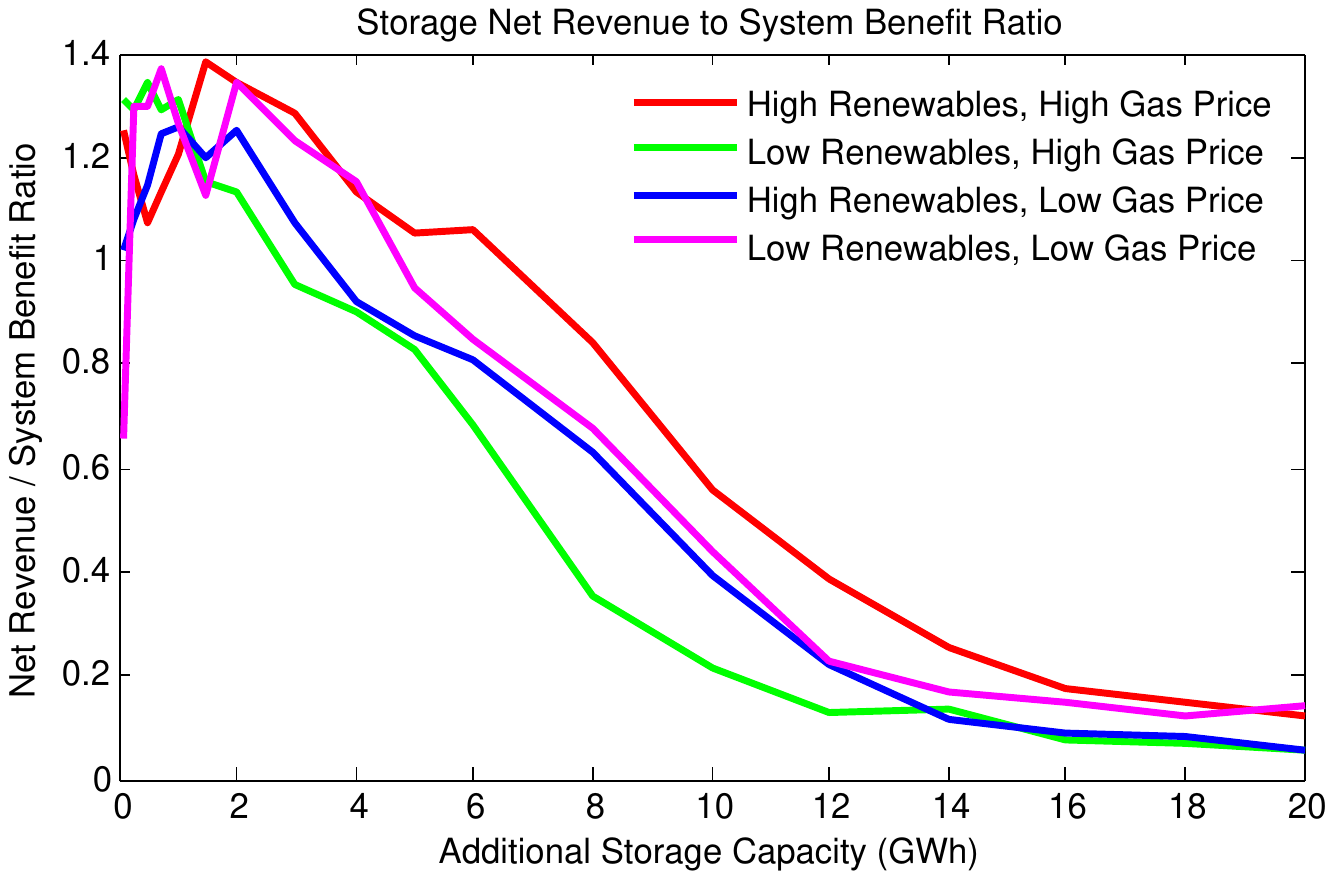}
\caption{Ratio of net revenue obtained by storage to system benefit provided by storage, relative to the base case with no storage. Between 4 and 6 GWh of storage the ratio dips below 1, which indicates that the system benefits provided by storage operationally are no longer able to be captured by storage operators via modeled markets. \label{fig:rev_ratio}}
\end{center}
\end {figure}

Figure~\ref{fig:rev_ratio} shows the ratio of the estimated market revenue ($Z$) for storage to its corresponding operating cost savings. In the figure, the ratio dips below one between 4 GWh and 6 GWh of storage capacity. This indicates that at this point, the system benefits from the presence of storage are no longer captured by storage operators through the markets modeled here. Primarily, these unaccounted for system benefits are realized as avoided starts when minimum up/down time constraints or ramp constraints would otherwise bind. Because storage operators are reliant on prices that are set by marginal costs of generators, storage operators are not able to realize the full value of their services, as the value of avoided starts is not reflected in the market clearing price for generation. 

\subsection{Carbon Emissions Due to Storage}

\begin{figure}
\begin{center}
\includegraphics[trim=7.1cm 9.2cm 7cm 10.8cm scale=0.90]{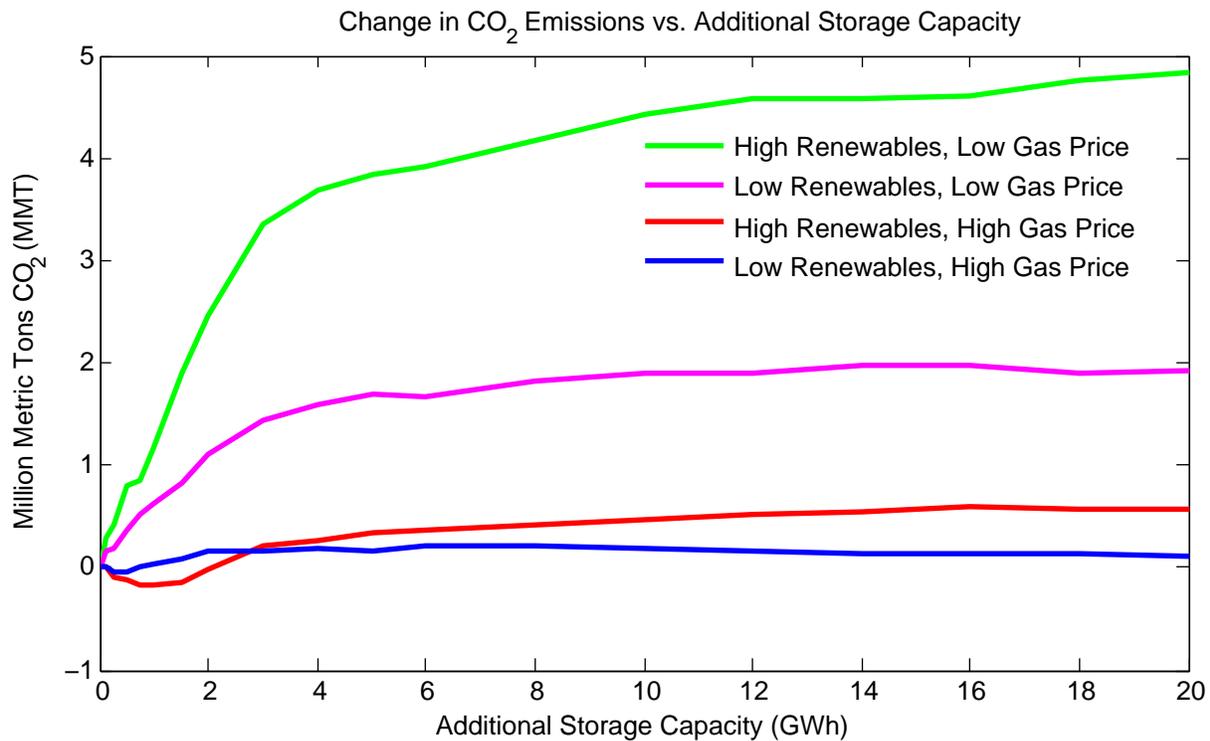}
\caption{As storage is added to the system, the carbon dioxide released due to system operations increases. In 2005, WECC emissions were between 370 and 385 MMT \cite{WECC_CO2}. \label{fig:CarbonEmissions}}
\end{center}
\end {figure}

\begin{figure}
\begin{center}
\includegraphics[trim=4cm 6cm 4cm 7cm]{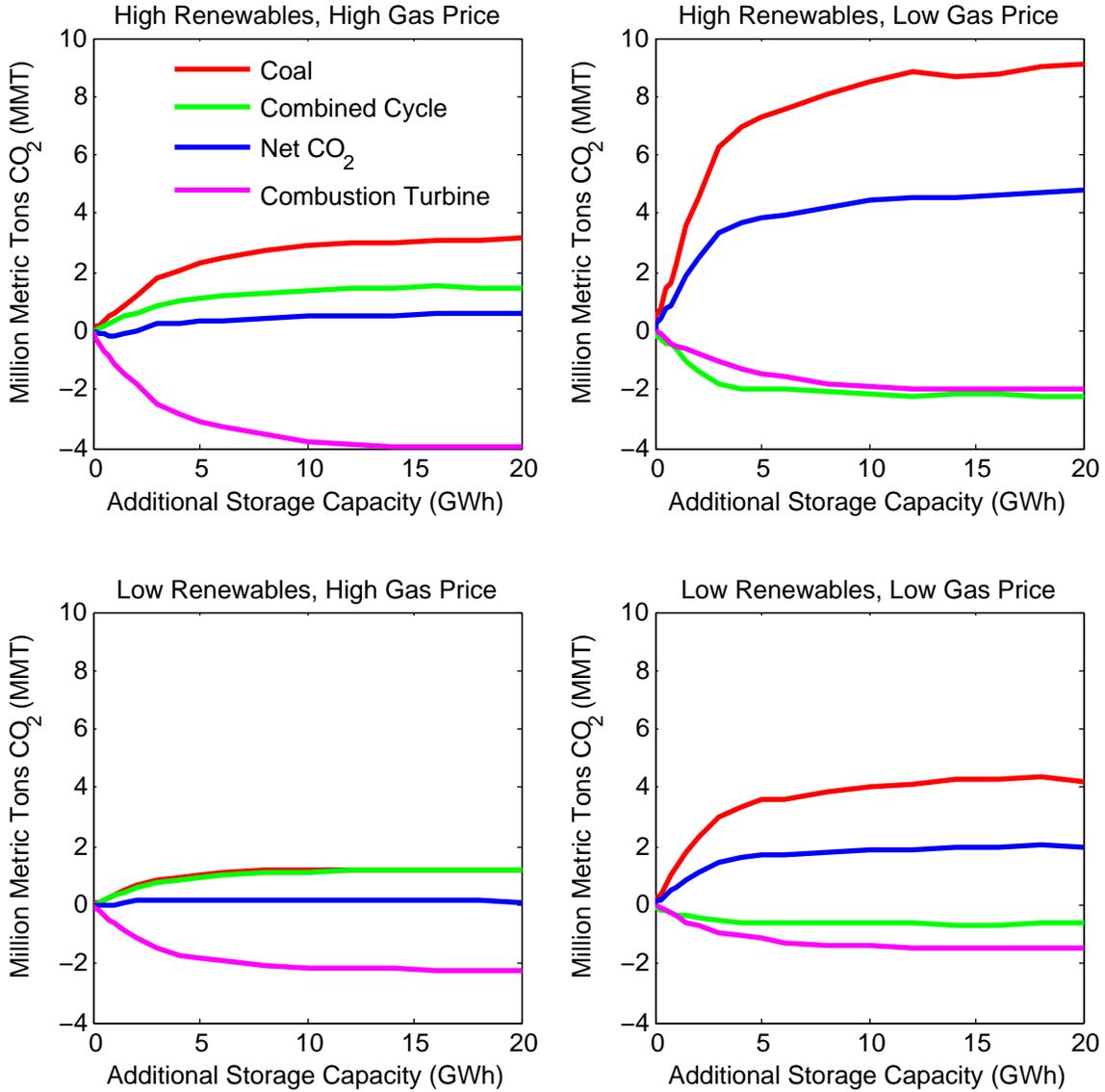}
\caption{This figure shows the emissions sources broken out by scenario. From Figure~\ref{fig:CarbonEmissions}, the largest increase in CO$_2$ emissions comes from a high renewables penetration and a low gas price.  In all scenarios, as the amount of storage added is increased, emissions from coal plants rise. In cases with a high gas price, emissions from combined cycle plants also rise, but with a low gas price they fall. In all scenarios, emissions from combustion turbines fall as the penetration of storage devices increases. \label{fig:CarbonEmissionsBreak}}
\end{center}
\end {figure}

As storage is added to the system, the carbon emissions associated with operating the system increase for most scenarios.  Figure~\ref{fig:CarbonEmissions} shows that carbon dioxide emissions strictly increase in most scenarios as storage penetration increases. 
In the high henewables, high gas price scenario carbon dioxide emissions experience a slight decrease until 1 GWh of storage capacity is present, and then emissions begin to increase. This effect is driven by fuel switching (from gas to coal), as shown in Figure~\ref{fig:CarbonEmissionsBreak}. 
When gas prices are low, there is less incentive to turn off gas plants to save money, since the savings from plant cycling is not enough to overcome large startup costs. This causes there to be more resources available on the system that can act as base load. The gas plants that are held on at their minimum capacities to avoid future startup costs are still more expensive to use on the margin than coal plants during operation, so coal plants are more frequently used at higher capacities. 

It is particularly notable that the most significant emissions increases happen in the high renewables / low gas price scenario, which may well be the most likely in light of energy futures prices and renewables capacity growth rates.  We will revisit this issue in the conclusions.

\section{Discussion}

We find that both fuel prices and renewables penetrations have a strong impact on the operational savings (see Figure~\ref{fig:savings_storage_capacity}), with fuel prices having a larger influence across the scenarios we investigated. 
This is driven in large part by the substantial difference in fuel prices in the 2007-2012 range we considered and particularly the large difference between the price of gas-fired peaker pants and coal-fired baseload plants (see the step change in marginal costs at 130 GW in the supply curve in Figure~\ref{fig:supply_curve}).  

Having high concentrations of renewables also corresponds to important operating cost benefits for storage --- savings from storage in high renewables situations is roughly double what it is in the low penetration scenarios we investigated.  This is due to the increased reserve requirements. As we showed in Figure~\ref{fig:ReserveBenefits}, in all scenarios the most valuable functions for storage to take over are reserve functions.  
We also note that operating cost benefits will further increase if one considers renewables penetrations beyond those we investigated; these benefits may eventually be comparable to the potential benefits at high gas prices.  However we do not observe large differences in the benefits from storage when the renewables mix is mostly solar rather than mostly wind, because it is optimal to use storage primarily for reserves, rather than arbitrage services.  This means that the timing of the resource is less important than its relative contribution to reserves requirements at the penetrations we investigated. 

The observation that it is more valuable for storage devices to provide reserves than arbitrage services is true from the perspectives of both storage operators and system operators. In all scenarios, regulation up is the most valuable service for storage to provide, followed by regulation down and load following up, and finally load following down. In general, load following provides the most revenue to storage operators, primarily because the market for load following is larger than the market for regulation. We also observe that the presence of storage has the potential to reduce both the total number and the overall cost of generator starts.  These results echo~\cite{harris2012unit}, who also studied the impact of storage on unit commitment and reserve provision (though not for a range of storage penetration levels and with a focus on CAES), and found that low penetrations of storage appear to be the most sensible in the short run.  

In combination, these factors indicate that storage is most beneficial in a system that has both large reserve requirements, as in the high renewables cases, and a large difference in marginal costs between low- and high-cost plants, as in the high gas price cases. While it is likely that renewables will encourage increases in reserve requirements in future systems, it is less likely that the spread in fossil fuel generation prices will stay large. With the recent decrease in natural gas prices due to hydraulic fracturing and horizontal drilling (``fracking''), the energy supply system has moved away from a price dichotomy that is advantageous for storage, and is closer to a system in which storage has a smaller effect on the economics of operations\footnote{Of course, though it may seem unlikely, future prices could change just as suddenly as they did with the introduction of fracking, and we cannot rule out a future fuel price scenario that favors more energy storage.}. In the scenarios with higher renewables penetrations and lower gas prices, the operating cost benefits achieved with storage are unlikely to justify capital cost expenditures on storage, even at aggressive capital cost estimates and low penetrations of storage. In these cases, it is very likely that the capacity value for storage will dominate any of the operational benefits we model here. 

With respect to carbon emissions, the presence of storage on the system causes an increase in CO$_2$ emissions for all scenarios, except at very small storage penetrations in the high renewables / high gas case. This is due to increased usage of coal plants in lower demand, low price hours to charge storage devices.  As long as the marginal price of electricity from gas exceeds that for coal (as it does in all scenarios we investigated), the cheapest times to charge storage devices will tend to be in hours when there are more coal plants on.  This implied that the energy stored in and then delivered by the storage devices will be dirtier than the energy supplied without storage. Relative to system-wide emissions, the increases are small; around a 1.4\% increase in emissions from the 2005 level.  It is worth noting other recent work has come to similar conclusions but with different modeling assumptions; in \cite{hittinger2015bulk}, the authors only examined bulk energy shifting, or arbitrage, but also found that storage increased carbon emissions.  

Overall, the benefits from increasing the presence of storage decline rapidly as installed capacity increases.  At 10 GWh of installed capacity, operating cost benefits for all but the high renewables / high gas price scenario are negligible.  In the high renewables / high gas case, benefits at 10 GWh are less than \$100/kWh; this is well below current prices but potentially achievable in the future if storage cost targets are met.  However in most scenarios carbon intensity continues to increase beyond 10 GWh of installed capacity.  This suggests that for the infrastructure we modeled --  a simplified version of WECC -- targets to install more than 10 GWh of energy storage are unlikely to be cost effective from an operating cost perspective in any near term future scenarios.  

As we noted above the capacity value of storage (approximately \$160/kWh assuming with 4 hour discharge capability storage could replace a \$650/kW combustion turbine) could become a very important part of its value as operating cost benefits decline.  If storage cost targets are met, capacity value alone could support significant expansion of storage capacity.  In this case we expect that the operating cost benefits we observe would still be realized, except on the limited peak net demand days when storage capacity is required for system reliability.

\section{Conclusions and Policy Implications}
Though the operational value of storage is high at very low penetrations, our analysis indicates that at modest penetrations (10 GWh, or 6 minutes of average energy demand in the model) the operational value is unlikely to compensate for storage capital costs in the foreseeable future.  To the extent storage is used to reduce operating costs, our analysis indicates that price arbitrage will be an insignificant factor, and that reserve provision will dominate.  This suggests that operating cost savings on their own do not likely constitute a motivation for policies that incentivize storage installations, even if it is on the expectation that those policies will indirectly drive installed costs downward.  However we note that reserve markets that capture the actual value of storage are in early stages; policy makers might consider initiatives to expand how much access storage owners have to reserve markets.   

Our analysis also indicates that operating cost savings quickly fall below plausible storage capacity values, and therefore capacity value is likely to be a significant component of the total transmission-level benefits of storage.  Indeed, if storage if cost targets are met, according to our calculations, generation capacity value alone justify the cost of storage.  However we made a simple assumption that 4 hours of energy storage capability would be sufficient to reproduce peaker plant capacity.  The total quantity of energy storage required for capacity value in practice could be more or less, depending on peak net load shapes but also the way storage is discharged in peak conditions.  Because it is energy-limited, operators will likely discharge storage conservatively to ensure system reliability.  If capacity value is to dominate operating cost benefits as a source of storage value, it will be important for storage owners, system operators, utilities and regulators to agree on best-practice discharge control algorithms in peak conditions.

We note, however, that our analysis did not investigate locational capacity value, both at the transmission level where ``load pockets'' can lead to very high local capacity costs and for  distribution systems where substation and conductor capacity may require upgrades to manage peak load growth.  These very important circumstances are beyond the scope of the present analysis; addressing them in detail would require detailed transmission models and circuit-level distribution capacity data.  

We found that storage operations can increase system-wide carbon emissions:  by reducing the required number of generator starts and providing flexible reserves, storage makes additional room for coal in the dispatch order.  It is important for regulators and system operators to consider policies and operating strategies that could be used to avoid this outcome.  However we expect that those policies would limit the operating cost benefits to storage and, as a consequence, diminish the financial incentive for storage owners to expand installed capacity.  



%

\section*{Acknowledgments}

This work was supported by the California Solar Initiative, a National Science Foundation Graduate Research Fellowship and NSF Grants 1351900 and 1239467.  We thank Anthony Papavasiliou for assistance with optimization algorithms, Jim Price at CAISO for providing the 240 bus WECC model, and members of the Energy Modeling, Analysis and Control Lab at UC Berkeley for guidance and input.  

\newpage

\bibliographystyle{elsarticle-harv}
\bibliography{StoragePaperBib2}

\end{document}